\documentclass[prd,aps,preprint,tightenlines,superscriptaddress,nofootinbib,showpacs,showkeys]{revtex4}
\usepackage{mathrsfs}
\usepackage{amsfonts}
\usepackage{array}
\usepackage{epsfig}
\usepackage{rotating}
\usepackage{multirow}

\usepackage{amsmath}    
\allowdisplaybreaks
\usepackage{verbatim}   
\usepackage{color}      
\usepackage{colordvi}
\usepackage{epstopdf}   

\usepackage{hyperref}   

\usepackage{pslatex}
\usepackage{txfonts}
\usepackage[latin1]{inputenc}
\usepackage[T1]{fontenc}

\def\mtmu{$m_t(\mu)$}
\def\mtmt{$m_t(m_t)$}
\def\mtpole{$m_t^{\rm pole}$}

\def\schannel{\mbox{$s$-channel}}
\def\tchannel{\mbox{$t$-channel}}
\def\ttbar{$t\bar t$}

\newcommand{\GeV}{\ensuremath{\,\mathrm{GeV}}}
\newcommand{\TeV}{\ensuremath{\,\mathrm{TeV}}}
\newcommand{\msbar}{$\overline{\mathrm{MS}}\, $}
\newcommand{\Hathor}{\textsc{HatHor}}

\begin{document}

\begin{titlepage}
\thispagestyle{empty}
\noindent
DESY 16-130\\
\hfill
August 2016 \\
\vspace{1.0cm}

\begin{center}
  {\bf
    \Large Determination of the top-quark mass \\[1ex]
    from hadro-production of single top-quarks}  \\
  \vspace{1.25cm}
  {\large
    S.~Alekhin$^{\, a,b}$,
    S.~Moch$^{\, a}$,
    and
    S.~Thier$^{\, a}$
   \\
 }
 \vspace{1.25cm}
 {\it
   $^a$ II. Institut f\"ur Theoretische Physik, Universit\"at Hamburg \\
   Luruper Chaussee 149, D--22761 Hamburg, Germany \\
   \vspace{0.2cm}
   $^b$ Institute for High Energy Physics \\
   142281 Protvino, Moscow region, Russia\\
 }
  \vspace{1.4cm}
  \large {\bf Abstract}
  \vspace{-0.2cm}
\end{center}
We present a new determination of the top-quark mass $m_t$ based on the
experimental data from the Tevatron and the LHC for single-top hadro-production. 
We use the inclusive cross sections of $s$- and \tchannel\ top-quark production to extract $m_t$ and to 
minimize the dependence on the strong coupling constant and the gluon distribution 
in the proton compared to the hadro-production of top-quark pairs. 
As part of our analysis we compute 
the next-to-next-to-leading order approximation for the \schannel\ cross section
in perturbative QCD based on the known soft-gluon corrections 
and implement it in the program \Hathor\ for
the numerical evaluation of the hadronic cross section.
Results for the top-quark mass are reported in the \msbar\ and in the on-shell
renormalization scheme.
\end{titlepage}

\newpage
\setcounter{footnote}{0}
\setcounter{page}{1}

Since the discovery of the top-quark in 1995~\cite{Abe:1995hr,Abachi:1995iq}, 
the precise value of its mass has always been of great interest 
as a fundamental parameter of the Standard Model (SM).
In the course of time several approaches have been used to extract the
top-quark mass $m_t$ as summarized for instance in~\cite{Agashe:2014kda}.
While kinematic fits to the top-quark decay products allow for a very precise determination 
of parameters in Monte Carlo (MC) programs that are used to describe the measured distributions, 
the relation of these MC parameters to the fundamental SM parameters 
needs to be calibrated and related uncertainties need to be taken into account~\cite{Kieseler:2015jzh}.
The determination of the top-quark mass from inclusive cross sections measured
at the hadron colliders Tevatron and the Large Hadron Collider (LHC) 
provides an alternative way. 
This allows to relate the experimental cross section measurements directly
to theoretical calculations which use a top-quark mass parameter 
in a well-defined renormalization scheme.

In this regard, the pair production of top-quarks has been of primary interest.
It is dominantly mediated by the strong interactions. 
In consequence, theoretical predictions for top-quark pair production are 
highly sensitive to the value of the strong coupling constant $\alpha_s$ as well as to the 
parton luminosity parameterized through the parton distribution functions (PDFs) of the colliding hadrons.
In fact, the uncertainty in the value of $\alpha_s$ and the dependence on the gluon PDF are the dominant sources 
which limit the precision of current theory predictions at the LHC~\cite{Accardi:2016ndt}.
Future measurements in particular at the LHC in Run 2 can potentially 
provide improved determinations of $\alpha_s$ and the PDFs, 
yet it is worth to investigate other methods to access $m_t$ 
that do not rely on these controversial quantities.

\begin{figure}[b!]
\begin{center}
\includegraphics[width=16.0cm]{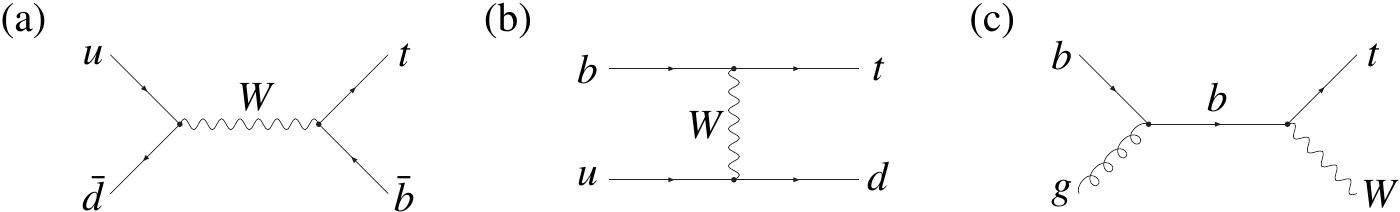}
\end{center}
\vspace*{-5mm}
\caption{\small
\label{fig:single_top_channels_diagrams}
Representative leading order Feynman diagrams for single top-quark production:
(a) \schannel; (b) \tchannel;  (c) in association with a $W$ boson.
}
\end{figure}

In this letter we determine the top-quark mass based on single-top production cross section measurements 
as a complementary way to arrive at a well-defined value for $m_t$ that is largely independent 
of $\alpha_s$ and the gluon PDFs.
Single-top production generates the top-quark in an electroweak interaction,
predominantly in a vertex with a bottom-quark and a $W$-boson.
The orientation of this vertex assigns single-top production diagrams to different
channels as illustrated in Fig.~\ref{fig:single_top_channels_diagrams}.
As our focus is on the minimization of the correlation between $m_t$, $\alpha_s$ and the gluon luminosity,
we consider only the so-called \schannel\ and \tchannel\ production 
of single top-quarks in the following. 
%
%
The cross sections for those processes are directly proportional to the 
light quark PDFs, which are nowadays well constrained by data 
on the measured charged lepton asymmetries from $W^\pm$ gauge-boson production
at the LHC. 
We use the inclusive single-top cross section measurements for those channels to
determine $m_t$ and compare the results to the ones obtained from \ttbar\ production.
Our study is based on data from the Tevatron at center-of-mass energy $\sqrt{S}=1.96~\TeV$ 
as well as from the LHC at $\sqrt{S}=7, 8$ and the most recent one at $13~\TeV$.

The theoretical description of both top-quark pair production
and single-top production has reached a very high level of accuracy.
The total cross section of \ttbar\ hadro-production 
has been calculated up to the next-to-next-to-leading order (NNLO)
corrections in perturbative QCD~\cite{Baernreuther:2012ws,Czakon:2012zr,Czakon:2012pz,Czakon:2013goa}.
The NNLO result shows good apparent convergence of the perturbative expansion  
and greatly reduced sensitivity with respect to a variation of the renormalization and
factorization scales $\mu_R$ and $\mu_F$, 
which is conventionally taken to estimate the uncertainty from the truncation 
of the perturbation series.

For the \tchannel\ of single-top production, the NNLO QCD corrections have been determined 
in the structure function approximation~\cite{Brucherseifer:2014ama} 
(see also Ref.~\cite{Berger:2016oht}),
by computing separately the QCD corrections to the light- and heavy-quark lines,
see Fig.~\ref{fig:single_top_channels_diagrams}~(b).
Any dynamical cross-talk between the two quark lines, e.g., double-box topologies, 
has been neglected in Ref.~\cite{Brucherseifer:2014ama} and  
is expected to be small due to color suppression.
The current theoretical status regarding those non-factorizing corrections is
summarized in Ref.~\cite{Assadsolimani:2014oga}.

The inclusive cross section of \schannel\ single-top production 
is fully known up to the next-to-leading order (NLO) QCD corrections~\cite{Smith:1996ij}, 
see also \cite{Harris:2002md} for fully differential results.
Beyond NLO accuracy, fixed-order expansions of the resummed soft-gluon contributions 
up to the next-to-leading logarithms (NLL) have been provided as an approximation 
of the complete NNLO result, both for the Tevatron~\cite{Kidonakis:2006bu} 
and the LHC~\cite{Kidonakis:2007ej}.
Subsequently, these result have been extended to next-to-next-to-leading logarithmic (NNLL) 
accuracy~\cite{Kidonakis:2010tc}.
The threshold corrections in the \schannel\ are large and dominant and,
therefore, they provide a good approximation to the full exact result, 
see Ref.~\cite{Alekhin:2016ioh} for a validation at NLO.
In our study we use Refs.~\cite{Kidonakis:2006bu,Kidonakis:2007ej,Kidonakis:2010tc}
to derive compact expressions for the approximate corrections at NLO and NNLO 
including soft-gluon effects almost complete to NNLL accuracy.
To that end, we integrate the partonic double-differential cross section 
given in Refs.~\cite{Kidonakis:2006bu,Kidonakis:2007ej,Kidonakis:2010tc}
over the phase space, i.e., the partonic Mandelstam variables $t$ and $u$,
and obtain the inclusive partonic cross section to logarithmic accuracy 
in the top-quark velocity $\beta = (1 - m_t^2/s)^{1/2}$.

We expand the partonic cross section for \schannel\ single-top production as a
power series 
\begin{equation}
\label{eq:sigma}
\sigma \,=\, \sigma^{(0)} + \alpha_s\, \sigma^{(1)} + \alpha_s^2\, \sigma^{(2)}\, ,
\end{equation}
with $\alpha_s=\alpha_s(\mu_R)$ taken at the renormalization scale $\mu_R$ 
and the leading-order partonic cross section for the process $u\bar{d} \to t\bar{b}$ given by 
\begin{equation}
\label{eq:sigma0}
\sigma^{(0)} =
\frac{\pi  \alpha^2 V_{tb}^2 V_{ud}^2 (m_t^2-s)^2 (m_t^2+2 s)}{24 s^2 \sin^4\theta_W (m_W^2-s)^2}\, .
\end{equation}
Here, $\sqrt{s}$ is the partonic center-of-mass energy, 
$m_W$ the $W$-boson mass and $\alpha$, $\sin\theta_W$, $V_{tb}$ and $V_{ud}$
are the electroweak and CKM parameters~\cite{Kant:2014oha}.

The NLO result in Eq.~(\ref{eq:sigma}) is denoted $\sigma^{(1)}$ 
and the exact result is known~\cite{Smith:1996ij} and has been implemented 
in the program \Hathor~\cite{Aliev:2010zk,Kant:2014oha} for a fast and
efficient evaluation of the total cross section.
Based on the threshold enhanced soft-gluon contributions 
we can provide an approximate NLO (aNLO) result for $\sigma^{(1)}$ as  
\begin{equation}
\label{eq:sig-aNLO}
\sigma^{(1)} \simeq 
\sigma^{(0)}\, \left(1-\beta^2\right)\, \frac{C_F}{8 \pi}\,  \left(
  112 \log^2(\beta )
  -148 \log (\beta )
  +63
  -4 \log\left(\frac{\mu_F^2}{m_t^2}\right) (8 \log (\beta )-3) 
\right)
+\mathcal{O}(\beta)\, ,
\end{equation}
where the coefficients of $\log^2(\beta)$ and $\log(\beta)$ are exact while we are 
lacking terms independent of $\beta$, i.e., ${\cal O}(\beta^0)$ from the
virtual contributions at one loop.
In addition we multiply the result by a kinematical suppression factor 
$(1-\beta^2) = m_t^2/s$ to restrict the soft-gluon logarithms to the threshold region.

The NNLO result $\sigma^{(2)}$ in Eq.~(\ref{eq:sigma}) is currently unknown,
but we can compute an approximate NNLO (aNNLO) expression for $\sigma^{(2)}$ 
valid near threshold $\beta \simeq 0$ as  
\begin{multline}
\label{eq:sig-aNNLO}
\sigma^{(2)} \simeq \sigma^{(0)}\, 
\left(1-\beta^2\right) \frac{C_F}{24 \pi ^2} \Bigg(
  2352 C_F \log^4(\beta )
  -8 \log ^3(\beta ) (17 \beta_0+777 C_F)
\\
  +\frac{1}{3} \log ^2(\beta ) \left(801 \beta_0 -28 \left(3 \pi ^2-67\right)
  C_A +24759 C_F -504 \pi ^2 C_F -280 n_f+\frac{144}{N_c}\right)
\\
  +\frac{1}{18} \log (\beta) \bigg(
    -4293 \beta_0 +C_A \left(3240 \zeta_{3}-18007+1008 \pi^2\right)
    +6480 C_F \zeta_{3}-111348 C_F
\\
    +4104 \pi ^2 C_F +2758 n_f-72 \pi ^2 n_f
    +\frac{3456}{N_c} \zeta_{3}+\frac{288}{N_c} \pi^2-\frac{7344}{N_c}
  \bigg)
\\
  -\frac{1}{120} \bigg(
    -10215 \beta_0 +25 C_A \left(648 \zeta_{3}-2315+144 \pi^2\right)
    +32400 C_F \zeta_{3}-251550 C_F
\\
    +11880 \pi ^2 C_F +8990 n_f-360 \pi ^2 n_f
    +\frac{23040}{N_c} \zeta_{3}+\frac{32}{N_c} \pi ^4
    +\frac{3840}{N_c} \pi^2-\frac{69120}{N_c}
  \bigg)
\\
  +\log \left(\frac{\mu_F^2}{m_t^2}\right) \bigg(
    -1344 C_F \log ^3(\beta )
    +12 \log ^2(\beta ) (7 \beta_0+190 C_F)
\\
    -\frac{1}{3} \log (\beta )
    \left(333 \beta_0-8 \left(3 \pi^2-67\right) C_A+6066 C_F-144\pi^2 C_F-80 n_f\right)
\\
    +\frac{1}{4} \left(189 \beta_0-8 \left(3\pi^2-67\right) C_A+3282 C_F-144 \pi^2 C_F-80 n_f\right)
\\
    +\log\left(\frac{\mu_R^2}{\mu_F^2}\right) (-24 \beta_0 \log (\beta )+18 \beta_0)
  \bigg)
\\
  +\log ^2\left(\frac{\mu_F^2}{m_t^2}\right) \left(
    192 C_F \log ^2(\beta )
    -12 \log(\beta )(\beta_0+12 C_F)
    +3 (3\beta_0+20 C_F)
  \right)
\\
  +\log\left(\frac{\mu_R^2}{\mu_F^2}\right) \left(
    84 \beta_0 \log ^2(\beta )-111\beta_0 \log (\beta )+\frac{189}{4} \beta_0
  \right)
\Bigg)
+\mathcal{O}(\beta) \hspace{25mm}
\end{multline}
where $\beta_0 = (11 C_A - 2 n_f)/3$ and $n_f$ is the number of quark flavors.
Moreover, we have $C_F=4/3$ and $C_A=3$ in QCD with $N_c=3$ colors and 
$\zeta_{3}$ denotes the Riemann $\zeta$-function.

In Eq.~(\ref{eq:sig-aNNLO}) all terms proportional 
to $\log^4(\beta)$ and $\log^3(\beta)$ are exact while 
those starting from $\log^2(\beta)$ are complete 
up to the interference of the one-loop threshold logarithms in Eq.~(\ref{eq:sig-aNLO})
with the ${\cal O}(\beta^0)$ part of the one-loop virtual corrections.
In our subsequent phenomenological studies we therefore restrict the use 
of threshold logarithms in Eq.~(\ref{eq:sig-aNNLO}).
For the scale independent part we keep all terms 
proportional to $\log^k(\beta)$ with $k=4,3,2$.
In analogy, we also keep the first three terms of the threshold expansion in Eq.~(\ref{eq:sig-aNNLO}) 
for all parts proportional to logarithms of $\mu_R$ or $\mu_F$,
that is $\log(\mu) \log^k(\beta)$ with $k=3,2,1$ 
and $\log^2(\mu) \log^k(\beta)$ with $k=2,1,0$.
In this way, we define the partonic cross section in the \schannel\ at approximate NNLO accuracy. 

\begin{figure}[!t]
\begin{center}
\includegraphics[width=0.85\textwidth]{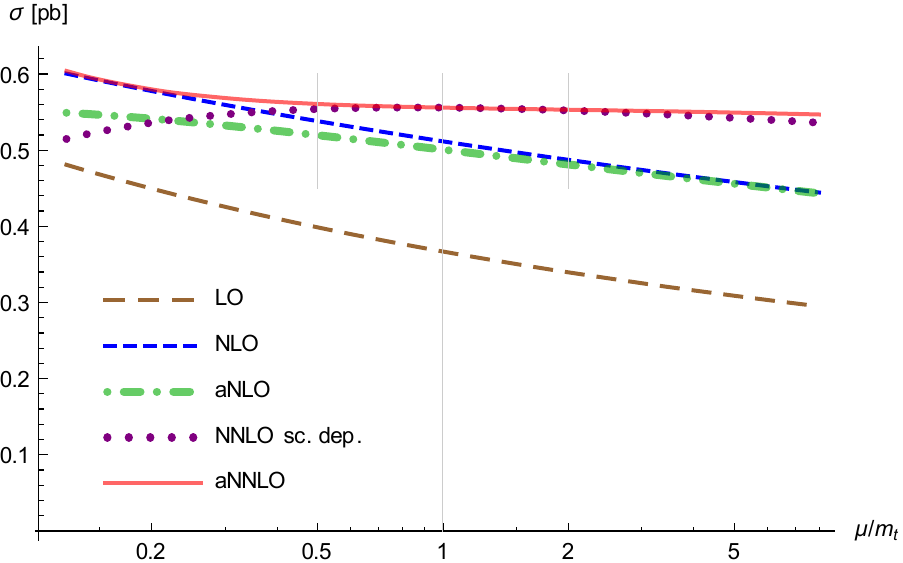}
\caption{\small
\label{pic:aNNLO-scale}
Cross section of single-top production in the \schannel\ 
for $p\bar{p}$ collisions using $\sqrt{S} = 1.96~\TeV$, 
\mtpole = 172.5 $\GeV$ and the ABM12 PDFs~\cite{Alekhin:2013nda} 
as function of $\mu/m_t$ with $\mu = \mu_R = \mu_F$ 
at LO (brown, long-dashed), 
at NLO (blue, short-dashed), 
at aNLO (green, dashed-dotted), 
at aNNLO (red, solid), 
and with scale dependence exact at NNLO (purple, dotted).
The vertical lines indicate the nominal scale $\mu = m_t$ 
and the conventional range $1/2 \le \mu/m_t \le 2$ for the variation.
}
\end{center}
\end{figure}

As a check of the convergence and the perturbative stability 
we show the scale dependence in Fig.~\ref{pic:aNNLO-scale} at LO, NLO and NNLO
for $p\bar{p}$ collisions at $\sqrt{S} = 1.96~\TeV$.
%
%
We focus here mainly on Tevatron kinematics for \schannel\ single-top production, 
since this process has not yet been established as an accurate enough observation at the LHC.
We use a pole mass \mtpole = $172.5~\GeV$, the PDFs of the ABM12 set~\cite{Alekhin:2013nda} 
and we identify $\mu=\mu_R=\mu_F$.
At NLO we plot the exact result~\cite{Smith:1996ij} and compare to the
threshold approximation for $\sigma^{(1)}$ given in Eq.~(\ref{eq:sig-aNLO}) 
and show that it approximates the exact result very well. 
In fact, around the nominal scale $\mu=m_t$ the deviations of the aNLO 
result Eq.~(\ref{eq:sig-aNLO}) from the exact one typically amount to only $5\%$ 
or less for collider energies in the range $\sqrt{S} = 1$ to $5~\TeV$.
At NNLO, we use the result for $\sigma^{(2)}$ in Eq.~(\ref{eq:sig-aNNLO}) 
including the scale dependent terms and subject to the truncation discussed above.
As an alternative, instead of those scale logarithms 
we can use the exact scale dependence at NNLO, which is provided by the program
\Hathor\ in numerical form, see~\cite{Kant:2014oha}.
Again, the differences between the two results are small except for very small values of $\mu$. 
In this case, numerically large but power suppressed contributions ${\cal O}(\beta)$ 
in the scale dependent part cause variations which remain 
uncanceled by the scale independent terms in Eq.~(\ref{eq:sig-aNNLO}).
In the conventionally chosen range $1/2 \le \mu/m_t \le 2$ for the scale
variations indicated by the vertical lines in Fig.~\ref{pic:aNNLO-scale}
any differences in the methodology to estimate the NNLO corrections 
are small so that we consider Eq.~(\ref{eq:sig-aNNLO}) 
restricted to the first three terms of the threshold expansion to provide a reliable 
approximation for the NNLO term $\sigma^{(2)}$ in Eq.~(\ref{eq:sigma}).
Below, we will use the residual scale dependence to estimate both 
the error due to the truncation of the perturbative expansion 
in Eq.~(\ref{eq:sigma}) as well as the systematic uncertainty 
inherent in the threshold approximation defining our aNNLO result. 
See also Ref.~\cite{Alekhin:2016ioh} for a further discussion of the validation 
of our approximation method.

The theoretical calculations for the hadro-production of top-quarks, singly or in pairs, 
typically use the on-shell renormalization scheme for the top-quark so that
the cross section predictions are given in terms of the pole mass \mtpole.
The advantages of other renormalization schemes which implement so-called short-distance masses, 
like the \msbar\ mass \mtmu\ at the scale $\mu$, have been discussed in the
literature at length, see for instance~\cite{Fleming:2007xt,Langenfeld:2009wd,Kieseler:2015jzh}.
The relation between the on-shell mass \mtpole\ and the \msbar\ mass is known up
to four loops in perturbation theory~\cite{Marquard:2015qpa,Marquard:2016dcn} 
and can be used to convert the respective cross sections. 
See for instance Refs.~\cite{Langenfeld:2009wd,Aliev:2010zk} 
for the derivation of $\sigma$(\mtmt) in terms of the \msbar\ mass 
\mtmt\ at $\mu=m_t$ from $\sigma$(\mtpole).
In summary, cross sections for the hadro-production of top-quark pairs exhibit 
a faster convergence and better scale stability if expressed in terms of the
\msbar\ mass.

This improved convergence is also observed for single-top production in the \schannel.
Evaluating the cross section for \schannel\ single-top production
in $p\bar{p}$ collisions at $\sqrt{S}=1.96~\TeV$ 
with the ABM12 PDFs and \mtpole = 172.5 $\GeV$, we find
$\sigma_{\text{LO}} = 0.37$~pb,
$\sigma_{\text{NLO}} = 0.51$~pb, and
$\sigma_{\text{aNNLO}} = 0.56$~pb,
which corresponds to an increase of 39\% at NLO relative to LO
and an increase of 9\% at aNNLO relative to NLO.
This growth is reduced when the cross section is
calculated for \mtmt = 163.0 $\GeV$.
In this case, we find the cross section values
$\sigma_{\text{LO}} = 0.47$~pb,
$\sigma_{\text{NLO}} = 0.57$~pb, and
$\sigma_{\text{aNNLO}} = 0.58$~pb
with an increase of 20\% at NLO relative to LO
and an increase of 3\% at aNNLO relative to NLO.

For the independent variation of both the renormalization scale $\mu_R$ 
and the factorization scale $\mu_F$ between
$\frac{1}{2}m_t$ and $2m_t$, excluding the points where
both scales are shifted in opposite directions, 
we see some increase in stability when using the \msbar\ mass.
In $p\bar{p}$ collisions at $\sqrt{S}=1.96~\TeV$ 
we find for a pole mass of 172.5~GeV variations
relative to the cross section at the central scale $m_t$
of +5.2\%/-4.7\% at NLO and +2.8\%/-2.4\% at aNNLO.
When the cross section is expressed as function of
the \msbar\ mass, which we set to 163~GeV here,
the scale dependence at NLO is reduced to
+3.1\%/-3.2\%. The scale dependence at aNNLO
is +3.6\%/-2.7\% for \mtmt, 
similar to though slightly larger than the scale dependence in the case of the pole mass.
The range of variations can be considered as an inherent uncertainty
of our approximation for Tevatron collisions.
At higher energies, like in $pp$ collisions at the LHC
with $\sqrt{S}=8~\TeV$, the threshold approximation is less accurate
and we find scale uncertainties of +5.3\%/-4.4\%
and +6.4\%/-5.3\% at aNNLO for the pole mass
and the \msbar\ mass respectively.

\begin{table}[t!]
\renewcommand{\arraystretch}{1.3}
\begin{center}                   
{\small                          
\begin{tabular}{|c|c|c|c|c|c|c|c|}   
\hline                           
Experiment                      
&\multicolumn{3}{c}{ATLAS}
&\multicolumn{3}{|c|}{CMS}  
&CDF \& D0
\\
\hline                                                    
{$\sqrt S$~(TeV)}                      
&{7}                         
&{8}                         
&{13}                         
&{7}                         
&{8}                         
&{13}                         
&{1.96}
\\                                                        
\hline
{Final states} 
& $tq$
& $tq$
& $tq$
& $tq$
& $tq$
& $tq$
&$tq, t\bar{b}$
\\
\hline                           
{Reference}                      
&\cite{Aad:2014fwa}                         
&\cite{Tepel:2014kna}
&\cite{Aaboud:2016ymp}
&\cite{Chatrchyan:2012ep}
&\cite{Khachatryan:2014iya}
&\cite{Sirunyan:2016cdg}
&\cite{Aaltonen:2015cra}
\\
\hline                                                    
{Luminosity (1/fb)}                      
&4.59                         
&20.3
&3.2
&2.73
&19.7
&2.3
&9.7x2
\\                                                        
\hline
{Cross section (pb)}                      
& $68 \pm 8$  
& $82.6 \pm 12.1$
&$247 \pm 46$
& $67.2 \pm 6.1$
& $83.6 \pm 7.7$
& $232 \pm 31 $
& $3.30^{+0.52}_{-0.40}$ (sum)
\\                                                        
\hline
\end{tabular}
}
\caption{
\label{tab:data-inp}
\small 
The data on single-top production 
in association with a light quark $q$ or $\bar{b}$-quark
from the LHC and Tevatron used in the present analysis. The errors given 
are combinations of the statistical, systematical, and  
luminosity ones.} 
\end{center}
\end{table}

Due to the pattern of improved convergence observed in all production processes, 
we use the \msbar\ scheme in our determination of the top-quark mass.
The fits to measured data are performed with the program \Hathor~\cite{Aliev:2010zk,Kant:2014oha}, 
which computes the inclusive cross sections for \ttbar\ and single-top production. 
In the \schannel, we implement our aNNLO result Eq.~(\ref{eq:sig-aNNLO})
for the partonic cross section in \Hathor\ 
and combine it with the built-in NLO formulae.
To evaluate the \tchannel\ total cross section, 
we use the NLO QCD predictions included in \Hathor\ and rescale them 
to account for the small NNLO QCD corrections calculated in Ref.~\cite{Brucherseifer:2014ama}.
%
%
In our analysis we use a common factor $k=0.984$ for the $t$ and $\bar{t}$ final states alike for this rescaling. 
This is justified as follows. 

For the \tchannel\ total cross section for a single $t$-quark Ref.~\cite{Brucherseifer:2014ama} 
reports a reduction by $-1.6\%$ at NNLO compared to NLO 
and for the one for a single $\bar{t}$-quark by $-1.3\%$, respectively.
Hence, there exists a slight dependence on the final state (see Tabs.~1 and~2 in Ref.~\cite{Brucherseifer:2014ama}).
It is worth poin-
%
%

%
\begin{sidewaystable}[ht!]
\renewcommand{\arraystretch}{1.3}
\begin{center}                   
{\small                          
\begin{tabular}{|c|c|c|c|c|c|c|c|c|}   
\hline                                                    
\multicolumn{2}{|c|}{}&                      
\multicolumn{6}{|c|}{Cross section (pb)}                      
\\
\hline                                                    
\multicolumn{2}{|c|}{$\sqrt S$~(TeV)}                      
&\multicolumn{2}{|c|}{7}
&\multicolumn{2}{|c|}{8}                         
&\multicolumn{2}{|c|}{13}                         
\\
\hline                                                    
\multicolumn{2}{|c|}{Experiment}                      
&{ATLAS}

&{CMS}  
&{ATLAS}
&{CMS}  
&{ATLAS}
&{CMS}  
\\                                                        
\hline
\multirow{2}{4em}{}
&dilepton + jets
& $181\pm11$~\cite{Aad:2014jra} 
&$174\pm6$~\cite{Khachatryan:2016mqs}
& 
&$245\pm9$~\cite{Khachatryan:2016mqs}
&
& $746\pm86$~\cite{Khachatryan:2015uqb}
\\
\cline{2-8}
\multirow{2}{*}{}
&dilepton + $b$-jet(s)
&$183\pm6$~\cite{Aad:2014kva}
&
& $242\pm9$~\cite{Aad:2014kva}
&
&$818\pm36$~\cite{Aaboud:2016pbd} 
& $793\pm44$~\cite{CMS:2016syx}
\\
\cline{2-8}
\multirow{2}{*}{Decay mode}
&lepton + jets
& 
&$162\pm14$~\cite{Khachatryan:2016yzq} 
&$260\pm24$~\cite{Aad:2015pga}
&$229\pm15$~\cite{Khachatryan:2016yzq}
&
& $836\pm133$~\cite{CMS:2015toa}
\\
\cline{2-8}
\multirow{2}{4em}{}
&lepton + jets, $b~\rightarrow~\mu~\nu~X$
&$165\pm17$~\cite{ATLAS:2012gpa} 
&
&
&  
&
& 
\\
\cline{2-8}
\multirow{2}{4em}{}
&lepton + $\tau\rightarrow$~hadrons 
&$183\pm25$~\cite{Aad:2015dya}
&$143\pm26$~\cite{Chatrchyan:2012vs}
& 
&$257\pm25$~\cite{Khachatryan:2014loa}
& 
&
\\
\cline{2-8}
\multirow{2}{4em}{}
&jets + $\tau\rightarrow$~hadrons
&$194\pm49$~\cite{Aad:2012vip} 
& $152\pm34$~\cite{Chatrchyan:2013kff}
& 
& 
&
& 
\\
\cline{2-8}
\multirow{2}{4em}{}
&all-jets
&$168\pm60$~\cite{ATLAS-CONF-2012-031}
& $139\pm28$~\cite{Chatrchyan:2013ual}
&  
& $276\pm39$~\cite{Khachatryan:2015fwh}
&
& $834^{+123}_{-109}$~\cite{CMS:2016rtp}
\\
\cline{2-8}                           
\hline
\end{tabular}
}
\caption{
\label{tab:data-tt}
\small 
The data on the $t\bar{t}$-production cross section
from the LHC used in the present analysis. The errors given 
are combinations of the statistical and systematical ones. 
An additional error of 3.3, 4.2 and 12~pb due to the beam energy uncertainty 
applies to all entries for the collision energy of $\sqrt{S}=7, 8$ and 13~TeV, respectively.
The quoted values are rounded for the purpose of a compact presentation. 
}
\end{center}
\end{sidewaystable}

\newpage
\cleardoublepage

%
%
\noindent
ting out, though, that the numbers reported in Ref.~\cite{Brucherseifer:2014ama}
implicitly depend on the perturbative accuracy of the chosen PDF sets as they have been obtained with a consistent use of PDFs, 
i.e. NLO (NNLO) PDFs for NLO (NNLO) predictions.
If we use NNLO PDF sets uniformly at every order for the cases considered in Ref.~\cite{Brucherseifer:2014ama}
we find a reduction of the cross section by $-1.2\%$ at NNLO compared to NLO, independent of the final state. 
This illustrates the limitations in accuracy of the rescaling method being at the level of a few per mill for the \tchannel\ total cross section, 
which is acceptable because any possible PDF dependence is small compared to the still sizable experimental uncertainties.

The cross section measurements of single-top production at the Tevatron and at the LHC 
that we use for our analysis are displayed in Tab.~\ref{tab:data-inp}.
%
%
For \schannel\ single-top production only Tevatron data are available.
In the \tchannel, we combine Tevatron data with the LHC ones at $\sqrt{S}=7$, 8 and 13 TeV. 
When a separation of $t$ and $\bar{t}$ final states 
is provided~\cite{Aad:2014fwa,Khachatryan:2014iya,Aaboud:2016ymp,Sirunyan:2016cdg}, 
we employ this information in our analysis. 
In this case a correlation between the systematic uncertainties in the 
single $t$- and $\bar{t}$-production data are taken into account 
using the error correlation coefficients 
\begin{equation}
  C_{t,\bar{t}} \,=\, \left(\delta \sigma_{t+\bar{t}}\right)^2 - \left(\delta \sigma_{t}\right)^2 - \left(\delta \sigma_{\bar{t}}\right)^2
  \, ,
  \label{eq:corr}
\end{equation}
where $\delta \sigma_{t}$, $\delta \sigma_{\bar{t}}$, and $\delta \sigma_{t+\bar{t}}$ are the 
systematic errors in the measured cross sections for the final states
containing a single $t$-quark, a $\bar{t}$-quark, and either $t$ or $\bar{t}$, respectively.  
The impact of the systematics correlation encoded in Eq.~(\ref{eq:corr}) turns
out to be more pronounced for the data samples of Refs.~\cite{Khachatryan:2014iya,Aaboud:2016ymp}
and it is marginal for the ones of Refs.~\cite{Aad:2014fwa,Sirunyan:2016cdg}.
Here, the luminosity errors quoted in Refs.~\cite{Aaboud:2016ymp,Sirunyan:2016cdg} are 
taken as fully correlated between the separated final states. 

We extract the $t$-quark mass also from data on $t\bar{t}$-production for comparison.
All inclusive cross sections obtained at the LHC at $\sqrt{S}=7$, 8, and 13 TeV are 
summarized in Tab.~\ref{tab:data-tt}.
These samples are categorized by the $t$-quark decay channels containing 
different numbers of the final-state leptons and jets. 
The systematic uncertainties in different channels and energies 
are taken as uncorrelated in general, however, the errors due to 
beam energy and luminosity 
are correlated for the data collected at the same 
collision energy. 
In addition to the data listed in Tab.~\ref{tab:data-tt} we also 
employ a combination of the measurements in different 
channels performed at Tevatron~\cite{Aaltonen:2013wca} and the 
recent CMS data~\cite{CMS:2016pqu} for the $e\mu$ decay
channel at $\sqrt{S}=5~\TeV$.

\begin{table}[t!]
\renewcommand{\arraystretch}{1.3}
\begin{center}                   
{\small                          
\begin{tabular}{|c|c|c|c|c|c|c|}   
\hline                           
Channel 
&ABM12~\cite{Alekhin:2013nda}
&ABMP15~\cite{Alekhin:2015cza}
&CT14~\cite{Dulat:2015mca}
&MMHT14~\cite{Harland-Lang:2014zoa}
&NNPDF3.0~\cite{Ball:2014uwa}
\\
\hline                                                    
\ttbar\
&$158.6\pm 0.6$                
&$158.4\pm 0.6$  
&$164.7\pm 0.6$
&$164.6\pm 0.6$
&$164.3\pm 0.6$
\\
\hline                                                    
\tchannel\
&$158.7\pm 3.7$                
&$158.0\pm 3.7$  
&$160.1\pm 3.8$
&$160.5\pm 3.8$
&$164.0\pm 3.8$
\\
\hline                                                    
$s$- \& \tchannel\
&$158.4\pm 3.3$                
&$157.7\pm 3.3$  
&$159.1\pm 3.4$
&$159.6\pm 3.4$
&$162.4\pm 3.5$
\\
\hline
\end{tabular}}
\end{center}
\caption{
\label{tab:mt-values-msbar}
\small 
Results for the running mass \mtmt\ in the \msbar\ scheme 
from the data listed in Tabs.~\ref{tab:data-inp} and \ref{tab:data-tt}
using different PDFs.} 
\end{table}

Our results for \mtmt\ from the fit to
single-top cross sections using the different
modern PDF sets 
  ABM12~\cite{Alekhin:2013nda}, 
  ABMP15~\cite{Alekhin:2015cza}, 
  CT14~\cite{Dulat:2015mca}, 
  MMHT14~\cite{Harland-Lang:2014zoa},  
  and 
  NNPDF3.0~\cite{Ball:2014uwa}
are collected in Tab.~\ref{tab:mt-values-msbar} together with corresponding mass values that are derived 
with the help of the \ttbar\ cross section data.
The uncertainties in Tab.~\ref{tab:mt-values-msbar} correspond to the ones which were
reported by the experiments for the respective data.
In addition, there are theoretical uncertainties $\Delta m_t$ 
from the variation of the factorization and renormalization scales in the
usual range $\frac{1}{2}$\mtmt$ \le \mu \le 2$\mtmt\ for $\mu=\mu_R=\mu_F$.
These are small and process dependent, but otherwise largely independent 
of the precise numerical value of the top-quark mass or 
of the specific PDF set considered in Tab.~\ref{tab:mt-values-msbar}.
We can quantify the effect of the scale variation on the extracted top-quark mass in the \msbar\ scheme 
as $\Delta m_t = \pm 0.7~\GeV$ for the \ttbar\ total cross section, see
e.g.~\cite{Alekhin:2013nda}.
Fits of the \msbar\ mass to Tevatron cross section data~\cite{CDF:2014uma}
for the respective scale choices
show mass uncertainties of $\Delta m_t = +0.9/-1.0~\GeV$
when our aNNLO approximation is used in the \schannel.
%
%
In that case we have to account for an additional $\Delta m_t = \pm 1.0~\GeV$ 
from the systematics of the threshold approximation used to define the aNNLO \schannel\ result.
The latter estimate is based on the accuracy of the threshold approximation at
NLO, i.e., the difference for the cross sections at the scale $\mu = m_t$
obtained either at NLO or at aNLO, cf. Fig.~\ref{pic:aNNLO-scale}.
For the \tchannel, we determine mass variations
at NLO accuracy in fits to the cross section
data that were reported in~\cite{Aaltonen:2015cra}
and subsequently take the reduced
scale dependence into account that was
found at NNLO~\cite{Brucherseifer:2014ama}.
In this way, we arrive at an uncertainty estimate
of $\Delta m_t = +0.6/-0.5~\GeV$ for our result in the \tchannel.

\begin{figure}[t!]
\begin{center}
\includegraphics[width=10.0cm]{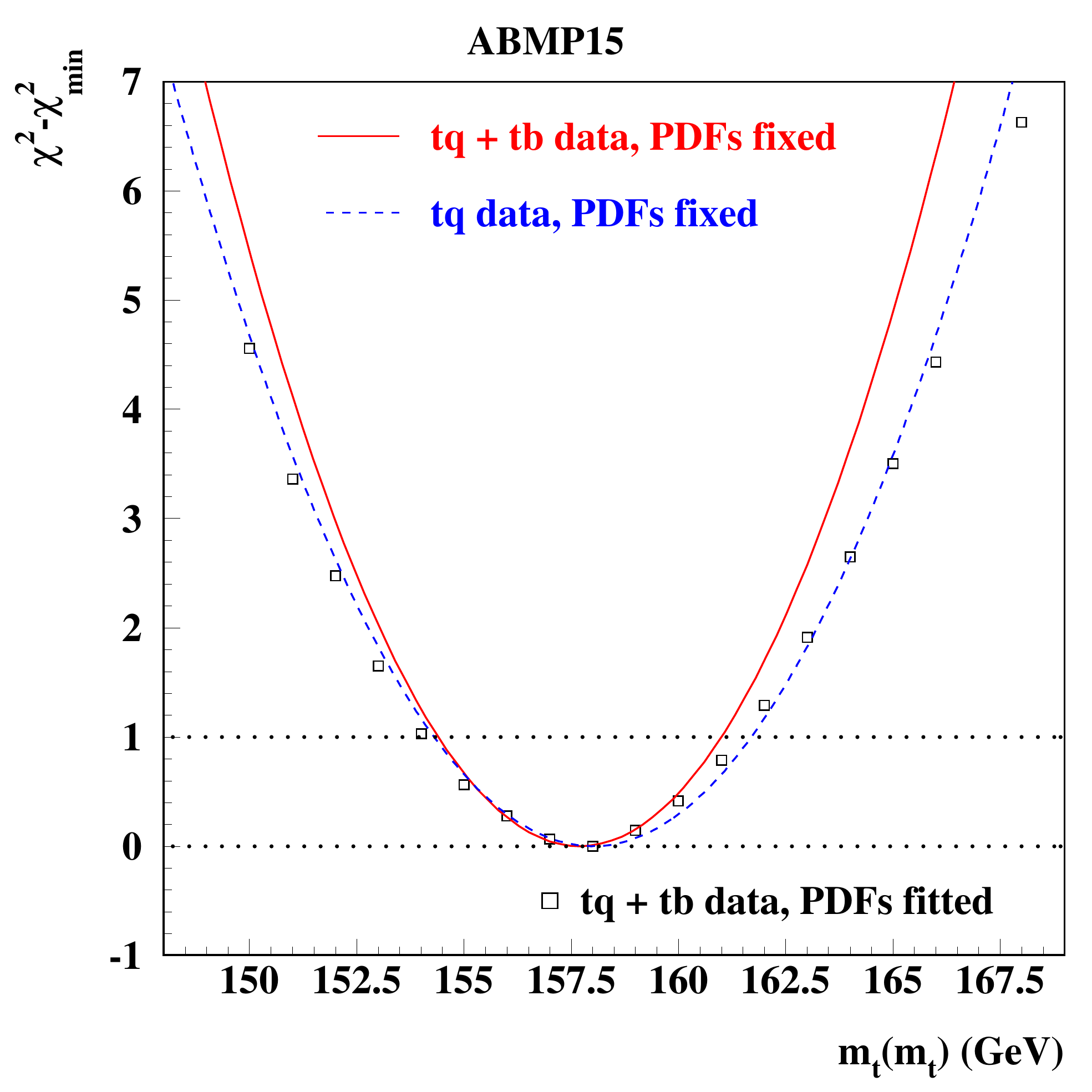}
\end{center}
\vspace*{-5mm}
\caption{
  \label{fig:scan}
  \small
  A profile of $\chi^2$ in a scan over the top-quark \msbar\ 
  mass obtained in the present analysis taking 
  the ABMP15 PDFs~\cite{Alekhin:2015cza} and the single-top production data 
  (solid: combination of the \schannel\ and \tchannel\ samples, 
  dashes: \tchannel\ sample only) 
  in comparison with the results obtained in the variant of the ABMP15 fit 
  with the \schannel\ and \tchannel\ single-top data appended (squares). 
  The minimal value $\chi^2_{\rm min} \sim 5$ is 
  subtracted in all cases.
}
\end{figure}

Due to the higher abundance of experimental data in the \tchannel, we report results of the mass fit
to either \tchannel\ data alone or the combination of all considered single-top data in $s$- and \tchannel.
The inclusion of the \schannel\ data favors a slightly smaller mass value
compared to the fit based on \tchannel\ data alone, 
cf. also the $\chi^2$ plot in Fig.~\ref{fig:scan}.

In order to facilitate the comparison of our results
for the top-quark mass to other studies of $m_t$,
for instance an earlier analysis performed in Ref.~\cite{Kant:2014oha},
we provide a conversion of the \msbar\ masses in Tab.~\ref{tab:mt-values-msbar}
to the respective pole mass values in Tab.~\ref{tab:mt-values-pole}.
The resulting pole mass \mtpole\ in the second line is obtained 
from a scheme transformation to NNLO accuracy, using the program {\textsc{RunDec}}~\cite{Chetyrkin:2000yt} 
and the value of $\alpha_s(m_Z)$ of the given PDF set.

\begin{table}[t!]
\renewcommand{\arraystretch}{1.3}
\begin{center}                   
{\small                          
\begin{tabular}{|c|c|c|c|c|c|c|}   
\hline                           
Channel 
&ABM12~\cite{Alekhin:2013nda}
&ABMP15~\cite{Alekhin:2015cza}
&CT14~\cite{Dulat:2015mca}
&MMHT14~\cite{Harland-Lang:2014zoa}
&NNPDF3.0~\cite{Ball:2014uwa}
\\
\hline                                                    
\ttbar\
&$167.3\pm 0.6$
&$167.1\pm 0.6$  
&$174.1\pm 0.6$
&$174.0\pm 0.6$
&$173.7\pm 0.6$
\\
&($167.9\pm 0.6$)
&($167.6\pm 0.6$)
&($174.7\pm 0.6$)
&($174.6\pm 0.6$)
&($174.3\pm 0.6$)
\\
\hline                                                    
\tchannel\
&$167.4\pm 3.9$                
&$166.7\pm 3.9$
&$169.3\pm 4.0$
&$169.7\pm 4.0$
&$173.4\pm 4.0$
\\
&($168.0\pm 3.9$)
&($167.2\pm 3.9$)
&($169.9\pm 4.0$)
&($170.3\pm 4.0$)
&($174.0\pm 4.0$)
\\
\hline                                                    
$s$- \& \tchannel\
&$167.1\pm 3.5$                
&$166.4\pm 3.5$
&$168.2\pm 3.6$
&$168.7\pm 3.6$
&$171.7\pm 3.7$
\\
&($167.6\pm 3.5$)
&($166.9\pm 3.5$)
&($168.8\pm 3.6$)
&($169.4\pm 3.6$)
&($172.3\pm 3.7$)
\\
\hline
\end{tabular}}
\end{center}
\caption{\small
  \label{tab:mt-values-pole}
  Results for \mtpole\ for different PDFs
  from the conversion of \mtmt\ at NNLO 
  (in parenthesis at N$^3$LO) using the 
  value of $\alpha_s(m_Z)$ corresponding to the respective PDF set.
} 
\end{table}

Interestingly, the results in Tabs.~\ref{tab:mt-values-msbar} and \ref{tab:mt-values-pole}
show a significant spread in the values of $m_t$ obtained for the different
PDF sets, but also when considering the different physical processes, i.e., the production of 
\ttbar-pairs versus single top-quarks in the $s$- and \tchannel.
For the PDF set ABM12 we obtain consistent values of \mtmt\ in Tab.~\ref{tab:mt-values-msbar}, 
i.e., central values of \mtmt$=158.6~\GeV$ from the \ttbar\ data 
and \mtmt$=158.4~\GeV$ from the combined $s$- and \tchannel\ data.
The results obtained for the ABMP15 set are very similar compared to those for ABM12.
The ABMP15 PDFs are based on an improved determination 
of the up- and down-quarks in the proton with the help of 
recent data on charged lepton asymmetries from $W^\pm$ gauge-boson production at the LHC and Tevatron. 
In particular, the ABMP15 PDFs find a non-zero iso-spin asymmetry of the sea, $x({\bar d}-{\bar u})$, 
at small values of Bjorken $x \simeq 10^{-4}$ and a delayed onset of the Regge asymptotics 
of a vanishing $x({\bar d}-{\bar u})$-asymmetry at small-$x$. 
This affects to some extent the cross section for \tchannel\ single-top production,  
but has overall little impact on the extracted value of \mtmt\ as can be seen from Tab.~\ref{tab:mt-values-msbar}. 

For the PDF sets CT14 and MMHT14 we find 
the central values \mtmt$=164.7~\GeV$ and \mtmt$=164.6~\GeV$ from the \ttbar\ data.
These are not only significantly larger than the ones obtained with 
ABM12 or ABMP15 due to the larger values for $\alpha_s(m_Z)$ and the gluon PDF in the relevant
$x$-range~\cite{Accardi:2016ndt}, 
but also much bigger than and barely compatible with the corresponding ones 
extracted from data for the single-top cross sections, 
\mtmt$=159.1~\GeV$ and \mtmt$=159.6~\GeV$.
This lack of compatibility at the level of 1$\sigma$ remains an issue even
when considering both the still sizeable uncertainty on \mtmt\ from the precision 
of experimental data as listed in Tab.~\ref{tab:mt-values-msbar}
and the theoretical uncertainty $\Delta m_t$ due the scale variation discussed above.
Finally, the \mtmt\ values determined with the NNPDF3.0 set 
are internally consistent yielding \mtmt$=164.3~\GeV$ and \mtmt$=162.4~\GeV$, respectively, 
when using the \ttbar\ data or the combined $s$- and \tchannel\ data.
However, they are significantly higher than the ones derived with the 
ABM12 and ABMP15 sets, so there is some tension among these two results.
All the observed differences are directly translated to the on-shell masses 
listed in Tab.~\ref{tab:mt-values-pole}.

Our study has shown that already with currently available data 
the top-quark mass can be determined to good accuracy 
for single-top cross sections and in doing so 
we have chosen the \msbar\ renormalization scheme for reasons of better perturbative stability.
The values obtained for the combined $s$- and \tchannel\ data can be used to
perform internal consistency checks for a given PDF set when comparing 
with the ones from \ttbar\ data.
Based on the dominant soft-gluon corrections 
we have provided new approximate predictions at NNLO for the inclusive \schannel\ single-top cross section 
and future theory improvements should complete the NNLO QCD correction to this process.
On the experimental side, high statistics measurements of single-top production 
at the LHC in Run 2 with $\sqrt{S}=13~\TeV$ can help substantially to  
further improve the precision of the top-quark mass.

\subsection*{Acknowledgments}
We would like to thank M.~Aldaya Martin for discussions.
This work has been supported by Deutsche Forschungsgemeinschaft in
Sonderforschungs\-be\-reich SFB 676.

{\footnotesize                                                          


}

\end{document}